\documentclass[aps,prd,preprintnumbers,nofootinbib,twocolumn]{revtex4}
\usepackage{bm}
\usepackage{latexsym}
\usepackage{dcolumn}
\usepackage{amsmath,amsfonts,amssymb}
\usepackage{graphicx,epsfig}
\usepackage{psfrag}
\usepackage{amsthm}

\def\be {\begin{equation}}
\def\ee {\end{equation}}
\def\bea {\begin{eqnarray}}
\def\eea {\end{eqnarray}}
\def\bc {\begin{center}}
\def\ec {\end{center}}
\def\bfg {\begin{figure}}
\def\efg {\end{figure}}
\def\bi {\begin{itemize}}
\def\ei {\end{itemize}}
\def\nn {\nonumber}

\def\la {\label}
\def\le {\left}
\def\ri {\right}
\def\pa {\partial}

\def\no {\noindent}

\def\vs {\vspace}

\def\a  {\alpha}

\def\beq{\begin{equation}}
\def\eeq{\end{equatfion}}
\def\br{\begin{eqnarray}}
\def\er{\end{eqnarray}}
\newcommand{\eel}[1] {\label{#1}\end{equation}}

\newcommand{\bdm}{\begin{displaymath}}
\newcommand{\edm}{\end{displaymath}}

\begin{document}
\title{Cosmology from quantum potential
}

\author{Ahmed Farag Ali $^{1,2}$}\email[email: ]{ahmed.ali@fsc.bu.edu.eg; afarag@zewailcity.edu.eg}
\author{Saurya Das $^3$} \email[email: ]{saurya.das@uleth.ca}

\affiliation{$^1$ Center for Theoretical Physics, Zewail City of Science and Technology, Giza, 12588, Egypt.\\}
\affiliation{$^2$ Dept. of Physics, Faculty of Sciences, Benha University, Benha, 13518, Egypt.\\}

\affiliation{$^3$ Department of Physics and Astronomy,
University of Lethbridge, 4401 University Drive,
Lethbridge, Alberta, Canada T1K 3M4 \\}

\begin{abstract}
It was shown recently that replacing classical geodesics with quantal (Bohmian)
trajectories gives rise to a quantum corrected Raychaudhuri equation (QRE).
In this article we derive the second order Friedmann equations from the QRE,
and show that this also contains a couple of quantum correction terms,
the first of which can be interpreted as cosmological constant
(and gives a correct estimate of its observed value),
while the second as a radiation term in the early universe,
which gets rid of the big-bang singularity and predicts an infinite age of our universe.
\end{abstract}

\maketitle


%
%
The generally accepted view of our universe
(homogeneous, isotropic, spatially flat,
obeying general relativity, and currently consisting of
about $72 \%$  Dark Energy, likely in the form of a
cosmological constant $\Lambda$, about $23\%$
Dark Matter, and the rest observable matter), implies its
small acceleration, as inferred from Type IA supernova observations,
CMBR data and baryon acoustic oscillations
\cite{perlmutter,riess,wmap,bao}.
However, quite a few things remain to be better understood, e.g., \\
(i) the smallness of $\Lambda$, about
$10^{-123}$ in Planck units ({\it `the smallness problem'}), \\
(ii) the approximate equality of vacuum and matter density in the current epoch
({\it `the coincidence problem'}), \\
(ii) the apparent extreme fine-tuning required in the early universe, to have a spatially
flat universe in the current epoch ({\it `the flatness problem'}), \\
(iv) the true nature of dark matter, and \\
(v) the beginning of our universe, or the so-called big-bang.

In this article, we show that one may be able to get
a better understanding of some of the above problems by studying the
quantum correction terms in the second order Friedmann equation, derived from the
quantum corrected Raychaudhuri equation (QRE), which in turn was
obtained by replacing geodesics with quantal (Bohmian) trajectories \cite{bohm}
(This formulation of quantum mechanics gives rise to identical predictions as those of ordinary quantum mechanics).
In particular, while one correction term can
be interpretable as dark energy, with the right density, and providing a possible
explanation of the coincidence problem,
the other term can be interpreted as a radiation term in the
early universe, preventing the formation of a big-bang type singularity, and predicting an infinite
age of our universe.
One naturally assumes a quantum mechanical description of the fluid or condensate filling our universe,
described by a wavefunction $\psi={\cal R}e^{iS}$
(assumed normalizable and single valued.
Some well-studied examples in curved spacetimes, including in cosmology, include
refs.\cite{hawking,melnikov,hartle,tumulka}.
${\cal R} (x^\a),S (x^a)=$ real functions), associated with
the four velocity field $u_a = (\hbar/m) \pa_a S$, and expansion
$\theta=Tr(u_{a;b}) = h^{ab} u_{a;b},~h_{ab}=g_{ab}-u_au_b$
(with vanishing shear and twist, for simplicity.
The constant $\epsilon_1=1/6$ for conformally invariant scalar fluid, but left arbitrary here.).
We will see later in this article that a condensate composed of gravitons with a
tiny mass is a natural candidate for this fluid.
Then the quantum corrected Raychaudhuri equation follows \cite{sd}
\footnote{We use the metric signature $(-,+,+,+)$ here, as opposed to $(+,-,-,-)$ in \cite{sd},
resulting in opposite sign of the $\hbar^2$ terms.}
\bea
\frac{d\theta}{d\lambda} =&& - \frac{1}{3}~\theta^2
- R_{cd} u^c u^d  \nn \\
&& +  \frac{\hbar^2}{m^2} h^{ab} \le( \frac{\Box {\cal R}}{\cal R} \ri)_{;a;b}
 + \frac{\epsilon_1 \hbar^2}{m^2} h^{ab} R_{;a;b},
\la{qre2a} 
\eea
Note that Eq.(\ref{qre2a}) follows directly the Klein-Gordon or Dirac equation
(or the Schr\"odinger equation for non-relativistic situations), and the quantum corrected
geodesic equation that follows from them \cite{sd}.
The second order Friedmann equation satisfied by
the scale factor $a(t)$ can be derived from the above, by replacing
$\theta = 3{\dot a}/{a}~,$ 
and
$R_{cd} u^c u^d \rightarrow \frac{4\pi G}{3} (\rho+3p) - \Lambda c^2/3,$
\cite{akrbook}
\footnote{
This procedure, as well the rest of the paper assumes large scale homogeneity and isotropy.
Even if there are small (perturbative) deviations from homogeneity, these can be absorbed in an
effective density $\rho$. Further these do not affect the dark energy content and
accelerated expansion of the universe \cite{fry,wald}.}
\footnote{
Note that in \cite{alberto1} too, the authors studied dark energy from the Bohmian mechanics
perspective, but originating in a scalar field with non-standard action.
Also, recently in \cite{he}, the authors used Bohmian mechanics in the context of
Wheeler-DeWitt equation, to explain inflation.}
(here the cosmological constant $\Lambda$ has dimensions of $1/(\mbox{length})^2$ as usual.)
\bea
\frac{\ddot a}{a} &&= - \frac{4\pi G}{3} \le( \rho + 3p \ri) + \frac{\Lambda c^2}{3} \nn  \\
&& + \frac{\hbar^2}{3 m^2} h^{ab} \le( \frac{\Box {\cal R}}{\cal R} \ri)_{;a;b}
+\frac{\epsilon_1\hbar^2}{m^2}h^{ab}R_{;a;b}~,
\la{frw1}
\eea
where the density $\rho$ includes visible and dark matter, and may also include
additional densities that arise in massive non-linear theories of gravity
\cite{massive1,massive2,massive3}.
The $\hbar^2$ terms in  Eqs.(\ref{qre2a}) and (\ref{frw1})
represent quantum corrections (the first of these is also known as quantum potential), which vanish
in the $\hbar\rightarrow 0$ limit, giving back
the classical Raychaudhuri and the Friedmann equations.
Note that these additional terms are not ad-hoc or hypothetical, but rather an
unavoidable consequence of a quantum description of the contents of our universe.
Also, since it is well known that Bohmian trajectories do not cross
\cite{holland,nocrossing},
it follows that even when $\theta$ (or $\dot a$) $\rightarrow -\infty$, the actual trajectories
(as opposed to geodesics) do not converge, and there is no counterpart of
geodesic incompleteness, or the classical singularity theorems, and singularities such as big bang or big
crunch are in fact avoided. This view is also supported by the quantum corrected geodesic deviation equation derived in \cite{sd}, which suggested that trajectories can never actually access infinite curvatures
\footnote{A similar conclusion was also arrived at by the frequency dependence of light paths
(`gravity's rainbow') picture in \cite{ahmed}
.}.
We will return to this issue later, and consider the first of these terms,
which naturally appears as a cosmological constant
\bea
\Lambda_Q = \frac{\hbar^2}{m^2 c^2} h^{ab} \le( \frac{\Box {\cal R}}{\cal R} \ri)_{;a;b}~.
\la{qlambda}
\eea
$\Lambda_Q$ depends on the amplitude ${\cal R}$ of
the wavefunction $\psi$, which we take to be the macroscopic ground state
of a condensate (more on the details of condensate in \cite{dasbhaduri}).
Its exact form is not important to our argument
however, except that it is non-zero and spread out
over the range of the observable universe. This follows from the requirement of causality;
even if matter exists beyond the horizon, it will have no effect on what is inside the
horizon, including the wavefunction. To estimate $\Lambda_Q$, one may assume a Gaussian form
$\psi \sim \exp(-r^2/L_0^2)$, or for one which results when an
interaction of strength $g$ is included in a scalar field theory,
such that $\psi=\psi_0\tanh(r/L_0 \sqrt{2}) ~(g>0)$ and $\psi=\sqrt{2}~\psi_0~{\mbox {sech}}(r/L_0)~(g<0)$
\cite{rogel}, it can be easily shown that $(\Box {\cal R}/{\cal R})_{;a;b} \approx 1/L_0^4$,
where $L_0$ is the characteristic length scale in the problem, which is of the order of the
Compton wavelength $L_0=h/mc$ \cite{wachter}, over which the wavefunction is non-vanishing.
This gives
\bea
\Lambda_Q = \frac{1}{L_0^2} =\le( \frac{m c}{h} \ri)^2 ~, \la{lambda1}
\eea
which has the correct sign as the observed cosmological constant.
Next to estimate its magnitude, we identify $L_0$ with the current linear dimension
of our observable universe, since anything outside it would not influence an accessible
wavefunction. With this,
$m$ can be regarded as the small mass of gravitons (or axions),
with gravity (or Coulomb field) following a Yukawa type of force law
$
F = - \frac{Gm_1 m_2}{r^2} \exp(-r/L_0).
$
Since gravity has not been tested beyond this length scale,
this interpretation is natural, and may in fact be unavoidable \cite{dasbhaduri}.
If one invokes periodic boundary conditions, this is also the mass of the lowest
Kaluza-Klein modes.
Substituting $L_0 = 1.4 \times 10^{26}~m$, one obtains
$m \approx 10^{-68}~kg$ or $10^{-32}~eV$,
quite consistent with the estimated bounds on graviton masses from various experiments \cite{graviton},
and also from theoretical considerations \cite{zwicky,mann,derham,majid}.
%
%
In other words, we interpret the quantum condensate as made up of these gravitons, and
described by a macroscopic wavefunction.
Finally, plugging in the above value of $L_0$ in Eq.(\ref{lambda1}), we get
\bea
\Lambda_Q && = 10^{-52}~(metre)^{-2} \\
&& = 10^{-123}~~(\text{in Planck units})~,
\eea
which indeed matches the observed value.
Also, since the size of the observable universe is about $c/H_0$, where
$H_0$ is the current value of the Hubble parameter \cite{weinberg}, one sees why
the above value of $\Lambda_Q$ numerically equals $H_0^2/c^2$
(which is $8\pi G/3 c^4 \times \rho_{crit}$, the critical density), offering a viable
explanation of
the coincidence problem. Note that the above also implies that
this equality will hold at all times during the evolution of the universe.
While the above relationship may have been known, here we provide a natural explanation for it,
in terms of quantum corrections to the Raychaudhuri equations via the wavefunction of the cosmic fluid.
This also brings out the essential quantum origin of the small cosmological constant.

%

One may also be tempted to interpret the quantum potential term as a perfect fluid
with $w_Q \equiv p_Q/\rho_Q=-1/3$ (such that $\rho_Q \propto a^{-3(1+w_Q)} \propto a^{-2}$).
However, a careful analysis using chi-squared techniques, of the luminosity distances versus
red-shifts of $580$ union 2.1 Supernovae data points, baryon acoustic oscillations,
Hubble and CM shift parameters suggest that
the corresponding density does not exceed 5\% of the current density of the universe
(at 95\% confidence level), and also
does not play any significant role in early or late epochs \cite{wali}. We therefore
do not consider that possibility.

Next, we consider the second correction term in Eq.(\ref{frw1}), which can be
written in terms of $H=\dot{a}/a$, and for one species of fluid, with $p/\rho=w$ as
\footnote{The following discussions, and in particular the conclusion following from Eq.(\ref{infinitetime}),
remain valid even if the first (cosmological constant) term
is retained.
}
\bea
&& \dot{H} = -\frac{3}{2} (1+w)H^2 +\frac{\epsilon_1\hbar^2}{m^2}h^{ab}R_{;a;b}~ \nn \\
&& = -\frac{3}{2} (1+w) H^2 \nn \\
&&
- \frac{\epsilon_1 \hbar^2}{m^2} 6 H^4 (1+w)\Bigg[6 (1+w)^2-\frac{81}{2}(1+w)+18\Bigg]~,
\la{fried2}
\eea
where in the last step we have plugged in the FRW metric.
When $\hbar =0$, integration yields
$H\rightarrow\infty$ in a finite time, signifying a big-bang type of singularity.
It is interesting to note that
$H^4$ proportional terms were also obtained from
(i) the trace anomaly of a conformal field theory
dual to a five-dimensional Schwarzschild-AdS geometry, and which is known as
holographic/conformal-anomaly Friedmann equation
\cite{Apostolopoulos:2008ru,Lidsey:2009xz},
(ii) correction to Raychaudhuri equation in cosmology derived in brane world scenarios
\cite{Maartens:2010ar}, and
(iii) derived in spacetime thermodynamics  and the generalized uncertainty principle of quantum gravity
\cite{Lidsey:2009xz}.
It would be interesting to investigate underlying connections between the above if any.

\begin{center}
\begin{figure}
\includegraphics[scale=0.25,angle=0]{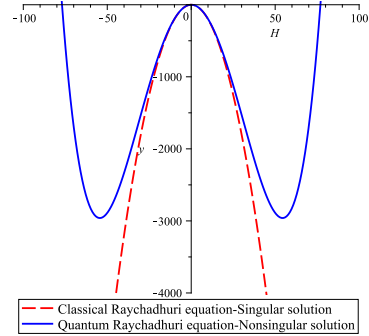}
\caption{$\dot{H}$ versus $H$}.
\label{solution}
\end{figure}
\end{center}

Next, to examine the presence or absence of past singularities, we write Eq.(\ref{fried2}) as
\bea
\dot{H}=F(H)~, \label{Conserv11}
\eea
and from which the age of the universe as
\bea
T = \int_0^T dt = \int_{H_0}^{H_P} \frac{dH}{F(H)}~,
\eea
where $H_0$ signifies the current epoch.
For the ordinary FRW universe with no quantum corrections of the above
type (i.e. $\epsilon_1=0$), $F(H)=-(3/2)(1+w^2)H^2$, the density
and $H \rightarrow \infty \equiv H_P$ in the past (big-bang singularity),
where $|\dot H| \rightarrow \infty$ as well, and we get
\bea
T = \frac{2}{3(1+w)^2 H_1}~,
\eea
which once again, is finite. Now if corrections to the classical FRW model changes the nature of the
function $F(H)$ (e.g. the degree of the polynomial), such that now neither $H$ nor
$\dot H$ diverges, then if
$H_P$ signifies the nearest fixed point in the past, such that $F(H_P)=0$, we
approximate $F(H) = F^{(n)}(H_P)(H-H_P)^n$ near the fixed point,
the region which contributes most to the integral, and obtain
\cite{Awad:2013tha}
\bea
T = \frac{1}{F^{(n)}(H_P)} \int_{H_1}^{H_P} \frac{dH}{(H-H_P)^n} \rightarrow \infty ~,
\label{infinitetime}
\eea
signifying a universe without a beginning. This is precisely what is expected from
the no-focusing of geodesics and the quantum Raychaudhuri equation.
For example, for Eq.(\ref{frw1}), it can be easily
shown that the sign of the quantum correction ($H^4$) term is positive (i.e. opposite to the
classical, $H^2$ term), for $-0.52 \leq w \leq 5.27$, which covers most of the physically
interesting range, including $w=w_R =1/3$ (radiation), which is most relevant for the very
early universe, and also non-relativistic matter ($w=w_{NR}=0$).
The situation is depicted in Figure (\ref{solution}), where it can be seen that
in the above range of $w$, $H_P$ is indeed finite, and therefore $T$ is infinite
from Eq.(\ref{infinitetime}). Thus, the second quantum correction in the Friedmann equation
gets rid of the big-bang singularity.

In summary, we have shown here that as for the QRE,
the second order Friedmann equation derived from the QRE also contains two quantum
correction terms.
These terms are generic and
unavoidable and follow naturally in a quantum mechanical description of our universe.
Of these, the first
can be interpreted as cosmological constant or dark energy of the correct (observed)
magnitude and a small mass of the graviton (or axion).
The second quantum correction term pushes back
the time singularity indefinitely, and predicts an everlasting universe.
While inhomogeneous or anisotropic perturbations are not expected to significantly affect
these results, it would be useful to redo the current study with such small perturbations to
rigorously confirm that this is indeed the case.
Also, as noted in the introduction, we assume it to follow general relativity,
whereas the Einstein equations may themselves undergo quantum corrections, especially at early epochs,
further affecting predictions. Given the robust set of starting assumptions,
we expect our main results to continue to hold even if and when a fully satisfactory
theory of quantum gravity is formulated.
For the cosmological constant problem at late times on the other hand,
quantum gravity effects are practically absent and can be safely ignored.
We hope to report on these and related issues elsewhere.

\vs{.2cm}
\no {\bf Acknowledgment}

\no
We thank S. Braunstein, M. W. Hossain, S. Kar, M. Sami, T. Sarkar and S. Shankaranarayanan for useful comments.
%
We thank the anonymous referee for useful comments.
This work is supported by the Natural Sciences and Engineering
Research Council of Canada, and the Perimeter Institute for Theoretical Physics,
through their affiliate program. The research of A. Farag Ali is supported by
Benha University (www.bu.edu.eg) and CFP in Zewail City, Egypt.
%



\end{document}